\documentclass[prc,pdftex]{revtex4}
\usepackage{graphicx,bm}
\usepackage{amsmath}

\begin{document}
\title{Effects of Momentum Conservation and Flow on Angular Correlations at RHIC}
\author{Scott Pratt and S\"{o}ren Schlichting}
\affiliation{Department of Physics and Astronomy and National Superconducting Cyclotron Laboratory,
Michigan State University\\
East Lansing, Michigan 48824}
\author{Sean Gavin}
\affiliation{Department of Physics, Wayne State University\\
Detroit, Michigan 48201}
\date{\today}

\begin{abstract}
Correlations of azimuthal angles observed at RHIC have gained great attention due to the prospect of identifying fluctuations of parity-odd regions in the field sector of QCD. Whereas the observable of interest related to parity fluctuations involves subtracting opposite-sign from same-sign correlations, the STAR collaboration reported the same-sign and opposite-sign correlations separately. It is shown here how momentum conservation combined with collective elliptic flow contributes significantly to this class of correlations, though not to the difference between the opposite- and same-sign observables. The effects are modeled with a crude simulation of a pion gas. Though the simulation reproduces the scale of the correlation, the centrality dependence is found to be sufficiently different in character to suggest additional considerations beyond those present in the pion gas simulation presented here.
\end{abstract}

\pacs{25.75.Gz,25.75.Ld}

\maketitle

\section{Introduction}\label{sec:intro}

When modeling heavy-ion collisions, one usually ignores momentum conservation. Given the thousands of hadrons emitted in central collisions at RHIC (the Relativistic Heavy Ion Collider), such effects are of the order of a fraction of a percent. However, as will be illustrated here, high-quality data sets measured by the STAR collaboration have sufficient statistics and coverage to identify and analyze correlations at levels better than $10^{-4}$, enabling the identification and analysis of correlations due to momentum conservation.

In the last year STAR has reported two-particle angular correlations of the form \cite{:2009uh,:2009txa},
\begin{equation}
\label{eq:gammadef}
\gamma\equiv \langle \cos(\phi_1+\phi_2)\rangle=\langle\cos\phi_1\cos\phi_2-\sin\phi_1\sin\phi_2\rangle,
\end{equation}
where $\phi_i$ refers to the azimuthal angle measured relative to the reaction plane. Correlations were reported for both same-sign, $\gamma_{\rm ss}$, and opposite-sign, $\gamma_{\rm os}$, pairs. The difference between the two correlations, $\gamma_{\rm qbal}\equiv \gamma_{\rm os}-\gamma_{\rm ss}$, was proposed to be a sign of parity fluctuations \cite{Kharzeev:2004ey,Kharzeev:2009mf}. The idea was that the parity-odd topological charge in the QCD sector would couple the large out-of-plane coherent magnetic fields from the passing ions to ${\bf E}_a\cdot{\bm B}_a$ in the QCD sector to create an out-of-plane electric field that would correlate out-of-plane charges of the same sign, or equivalently anti-correlate opposite-sign charges emitted out of plane. This would motivate a positive correlation for $\langle\sin\phi_1\sin\phi_2\rangle_{\rm ss}-\langle\sin\phi_1\sin\phi_2\rangle_{\rm os}$. To reduce the contributions from resonances, the $\cos\phi_1\cos\phi_2$ correlations were then subtracted from the $\sin\phi_1\sin\phi_2$ correlations, i.e., one would look for $\gamma_{\rm qbal}>0$. To better determine whether the observed angular correlations were evidence of large parity fluctuations, several more differential analyses have been performed: looking at the contributions to in-plane and out-of-plane separately \cite{Bzdak:2009fc,Wang:2009kd} or looking a charge balance functions \cite{soeren_bf,Schlichting:2010na}. Such studies suggest that most of the correlation in $\gamma_{\rm qbal}$ derives from another source. Upon further consideration it appears that most of the correlations in $\gamma_{\rm qbal}$ derives from charge conservation overlaid onto elliptic flow \cite{soeren_bf}. In fact, the more differential correlations encoded in charge balance functions \cite{Aggarwal:2010ya} were even reproduced at the quantitative level by overlaying local charge conservation onto a thermal blast-wave model which was tuned to single-particle spectra and elliptic flow.

The effects of charge conservation overlaid onto elliptic flow only addressed $\gamma_{\rm qbal}$ and are irrelevant for explaining why the same-sign correlation, $\gamma_{\rm ss}$, was fairly large and negative. Like the values for $\gamma_{\rm qbal}$, the magnitude of the correlation was found to be of the order $v_2/M$, where $M$ is the multiplicity and $v_2=\langle\cos 2\phi\rangle$ is the elliptic flow. It was argued in \cite{Pratt:2010gy} and \cite{Bzdak:2010fd} that such correlations might ensue from momentum conservation combined with elliptic flow. If for each particle emitted at angle $\phi_1$, there was a particle balancing the transverse momentum at angle $\phi_2=\phi_1-\pi$, one would find correlations, $\gamma\sim -v_2/M$.

The goals of this paper are: (1) to expand the discussions of momentum conservation in \cite{Pratt:2010gy} and to discern what combination of $\gamma_{\rm ss}$ and $\gamma_{\rm os}$, and with what weighting in momentum, would best isolate the effects of momentum conservation. (2) to demonstrate the effects of momentum conservation with a simulation, compare to experiment, and see whether the observed correlations are consistent with expectations from momentum conservation. (3) Investigate whether correlations of the class described above can be used to discern fundamental properties of the matter created at RHIC, e.g., the viscosity. The next sections presents the principles involved in observing momentum-conservation driven correlations with a focus on the momentum sum rule which drives the phenomenology. Section \ref{sec:vdiffusion} discuses the role of viscosity in spreading the correlation in relative rapidity. The pion cascade model is presented in Sec. \ref{sec:model} and compared to experimental results from STAR, while a summary and outlook are given in the final section.

\section{The Momentum Sum Rule}\label{sec:theory}

In the absence of competing correlations, such as charge conservation, the correlation $\langle \cos(\phi_1+\phi_2)\rangle$, averaged over all pairs regardless of charge, would be well suited to study momentum correlations. Even better would be the correlation, $\langle p_{x1}p_{x2}-p_{y1}p_{y2}\rangle/\langle p_t^2\rangle$, which is similar but is determined more directly by momentum conservation. Unfortunately, other correlations come into play which makes the choice of observable tricky. For instance, one might consider same-sign correlations as a way to avoid the effects of charge conservation. However, when one sees a positive particle at $\phi_1$, there is an enhancement for having a negative particle at a similar angle and with similar momentum. The momentum of the negative particle also needs to be balanced, and the balancing momentum could be shared by other positive particles. One must also consider the fraction of the balancing momentum carried by unobserved particles. For instance, in the STAR analysis neutral particles are not measured and even with the large acceptance and relatively high efficiency a good fraction of the balancing momentum is outside detection.

To better clarify the issues above, the theory discussion is split into parts. In the first subsection the situation will be considered for a system with a single species. The momentum sum rule will be derived for perfect acceptance and efficiency, with a discussion of how it would be modified for less than perfect detectors. The more realistic case, accounting for undetected neutral particles and the effects of charge conservation, is then considered in the following subsection. The relations found here are applied and tested with a simple model in Sec. \ref{sec:model}.

\subsection{The Idealized Case with No Other Correlations}

For illustrative purposes we consider the case of $M$ particles emitted under the constraint of momentum conservation. Rather than calculating $\gamma$ as defined in Eq. (\ref{eq:gammadef}), we consider the $p_t$ weighted correlation,
\begin{equation}
\label{eq:gammaprimedef}
\gamma'\equiv\frac{\langle p_{x1}p_{x2}-p_{y1}p_{y2}\rangle}{\langle p_t^2\rangle}.
\end{equation}
If all the particles had the same $p_t$, this would be no different than $\gamma$, which is defined solely through the angles. Although the angular correlations are stronger for higher $p_t$, $\gamma'$ surprisingly turned out to be several tens of percent smaller than $\gamma$ in the simulations of the next section. This is due to the denominator $\langle p_t^2\rangle$, which weights high-momentum particles even more strongly in the denominator than the extra $p_t$ weightings in the numerator. If the denominator had been defined as $\langle p_t\rangle^2$ rather than $\langle p_t^2\rangle$, $\gamma'$ would have been larger. The particular form in the definition of Eq. (\ref{eq:gammaprimedef}) is chosen to simplify the momentum sum rule shown below. Since this form is most directly related to the sum rule, it exhibits less model dependence than other choices.

To derive the momentum sum rule, one expands the definition of the averages assuming for the moment that all particles are measured,
\begin{eqnarray}
\gamma'&=&\frac{1}{M-1}\frac{\sum_{i\ne j}\left(p_{xi}p_{xj}-p_{yi}p_{yj}\right)}{\sum_i p_{ti}^2}\\
\nonumber
&=&\frac{1}{M-1}\frac{\sum_{i,j}\left(p_{xi}p_{xj}-p_{yi}p_{yj}\right)}{\sum_i p_{ti}^2}\\
\nonumber
&&-\frac{1}{M-1}\frac{\sum_{i=j}\left(p_{xi}p_{xj}-p_{yi}p_{yj}\right)}{\sum_i p_{ti}^2}\\
\nonumber
&=&\frac{1}{M-1}\frac{\sum_i p_{xi}^2-p_{yi}^2}{\sum_i p_t^2}=\frac{1}{M-1}\frac{\sum_i p_{ti}^2\cos 2\phi_i}{\sum_i p_t^2}.
\end{eqnarray}
Momentum conservation, $\sum_i p_{xi}=0$, was applied in the third line. The final term resembles an expression for $v_2$ weighted by $p_t^2$. Assuming $M$ is large one can derive the momentum sum rule,
\begin{equation}
\label{eq:v2primedef}
\gamma'=-\frac{1}{M}v_2',~~v_2'\equiv\frac{\langle p_t^2\cos 2\phi\rangle}{\langle p_t^2\rangle}.
\end{equation}

The sum-rule above is based on a perfect acceptance. Even for the STAR TPC detector, which has a broad acceptance of approximately $\pm 1$ units of rapidity, only about one third of the emission is addressed. Since the averages $\langle\cdots\rangle$ are correlations, the sum rule is not affected by a less than perfect efficiency if the efficiency is uniform. However, if the experiment has a limited acceptance, the sum rule can be significantly compromised. First of all, one must state whether $M$ refers to the multiplicity inside the acceptance, to the total multiplicity, or something in between. For instance, the STAR experiment has good coverage for pseudo-rapidities between -1 and 1 and for charged particles with transverse momenta above $\sim$200 MeV/$c$. One might choose $M$ as the multiplicity of particles with $p_t>200$ MeV/$c$, or one might use the usually reported measure of the multiplicity which assumes no $p_t$ cuts. If $M$ refers to the multiplicity within the acceptance, the sum-rule is modified by a factor $f_p$, where $f_p$ is defined as the fraction of balanced momentum which falls within the acceptance,
\begin{equation}
\label{eq:sumrule1species}
\sum_{j\ne i} p_{xj}=-f_P p_{xi}.
\end{equation}
Here, $f_P\le 1$, and the sum rule becomes,
\begin{equation}
\gamma'=-\frac{f_P}{M}v_2'.
\end{equation}
Part of the reduction, $f_P$, comes from balancing momentum that is carried by particles with rapidities outside the acceptance. This can be estimated by looking at $\gamma'$ as a function of relative rapidity and then convoluting that with the rapidity distribution. Given that the efficiency inside the acceptance will vary, the surest way to proceed is to model the correlations and simulate their measurement. Unfortunately, either strategy introduces some model dependence, which is unavoidable once acceptances become complicated.

Although transverse momentum is conserved globally, the balancing momentum for a specific particle might be, or might not be, spread over the entire collision volume. Whereas STAR's analysis cover approximately $\pm 1$ units of rapidity, emission is spread over a half dozen units of rapidity. In the thermal calculations by \cite{Bzdak:2010fd}, transverse momentum conservation was enforced globally. If one were to enforce conservation locally, the question would be ``How local?''. Clearly, addressing this question is central to estimating the fraction $f_P$ in Eq. (\ref{eq:sumrule1species}). During the initial interactions of the incoming nuclei, hard scatterings and fragmentation processes spread transverse momentum over a wide range of rapidity, whereas according to the commonly accepted picture of a RHIC collision, the subsequent processes are more local in nature, as the momentum transfers from subsequent collisions are shared amongst particles with similar rapidities. Further, it is these collisions that build up elliptic flow. Here, we will show how the lack of local momentum balance in the initial scattering does not affect the locality of Eq. (\ref{eq:sumrule1species}) as long as the particles created in the initial state have their momenta distributed isotropically in azimuthal angle.

First, one considers a distribution of particles with initial momentum $k_i$, whose final momentum is $p_i$ and whose net momentum transfer is $q_i$,
\begin{equation}
\vec{p}_i=\vec{k}_i+\vec{q}_i.
\end{equation}
We will assume that the vectors $\vec{k}_i$ are isotropic, i.e.,
\begin{equation}
\langle k_{xi}k_{xj}-k_{yi}k_{yj}\rangle = \langle k_{xi}^2-k_{yi}^2\rangle=0,
\end{equation}
and that the momentum transfers conserve momentum within the neighborhood,
\begin{equation}
\left\langle \sum_{j\ne i}\vec{q}_{j}\right\rangle=-f_P\vec{q_i}.
\end{equation}
Here, $f_P$ is again the fraction of the balancing momentum exchanges, $q_i$, that will be found within the acceptance. In the limit that the acceptance covers much more phase space than what is subtended by the diffusion of momentum from the transfers $q_i$, $f_P\rightarrow 1$, even if the collision then covers many more units. Calculating the azimuthal correlation,
\begin{eqnarray}
\label{eq:gammakqderiv}
\gamma'&=&\frac{1}{M(M-1)}\sum_{i\ne j}\left(
k_{xi}k_{xj}-k_{yi}k_{yj}+2k_{xi}q_{xj}\right.\\
\nonumber
&&\left.-2k_{yi}q_{yj}+q_{xi}q_{xj}-q_{yi}q_{yj}
\right)\\
\nonumber
&=&\frac{1}{M(M-1)}\sum_{i\ne j}\left(
2k_{xi}q_{xj}-2k_{yi}q_{yj}+q_{xi}q_{xj}-q_{yi}q_{yj}
\right)\\
\nonumber
&=&\frac{-f_P}{M(M-1)}\sum_i\left(
2k_{xi}q_{xi}-2k_{yi}q_{yi}+q_{xi}q_{xi}-q_{yi}q_{yi}\right)\\
\nonumber
&=&\frac{-f_P}{M}\langle 2k_xq_x-2k_yq_y+q_x^2-q_y^2\rangle,
\end{eqnarray}
where the isotropy of $\vec{k}$ and local conservation of $\vec{q}$ were both exploited. For the purposes of this derivation we consider a class of events with fixed $M$, but will take the limit of large $M$. 

One can also write an expression for the numerator for $v'_2$,
\begin{eqnarray}
\label{eq:v2kqderiv}
v^{\prime(\rm num)}_2&=&\langle k_x^2-k_y^2+2k_xq_x-2k_yq_y+q_x^2-q_y^2\rangle\\
\nonumber
&=&\langle 2k_xq_x-2k_yq_y+q_x^2-q_y^2\rangle,
\end{eqnarray}
where the isotropy of $\vec{k}$ was again used.

The equivalence of the the last expressions in Eq.s (\ref{eq:gammakqderiv}) and (\ref{eq:v2kqderiv}) confirms Eq. (\ref{eq:sumrule1species}). If the momentum transfers responsible for elliptic flow build up during the first few fm/$c$ of the reaction (but not in the initial state), one would expect the extent of $\gamma'$ as a function of relativity to depend on the diffusion of transverse momentum along the longitudinal direction, and thus perhaps on the shear viscosity. This will be explored in the next section.

\subsection{Including the Effects of Charge Conservation}

Charge conservation affects both same-sign and opposite-sign correlations. It is straight-forward to understand how opposite-sign correlations become manifest. Here, we first describe how charge conservation also affects same-sign correlations. Due to local charge conservation, each charge (aside from those carried by the incoming nuclei) is accompanied by an extra charge of the opposite sign. Since charge conservation is local, the two balancing charges will remain close to one another in coordinate space. The two balancing charges then flow together and are positively correlated in $\phi$. If one triggers on a specific positive particle, one expects momentum conservation to provide negative correlations in azimuthal angle, and that the correlation would be spread out roughly equally among positive, negative and neutral particles. The same-sign correlations would be expected to account for one third of the momentum balance. However, that balance is increased by the fact that the original positive particle is accompanied by a negative particle with similar momentum. This accompanying particle's momentum must also be balanced, and if it has the same momentum as the original triggered particle, the momentum balance one would find amongst other positive particles would double, i.e., one would expect nearly 2/3 of the original momentum of a triggered positive particle to be balanced by other positive particles. 

The first goal of this section is to find a combination of $\gamma'_{\rm ss}$ and $\gamma'_{\rm os}$ that leads to a momentum sum-rule despite the correlation inherent due to each positive being balanced by a negative charge, and the complication due to roughly one third of the particles being neutral, and in the case of STAR, undetected. To proceed we consider a system with multiplicities $M_{\pm}$ for both positive and negative particles and multiplicity $M_0$ of neutrals. Assuming that charged particles correlate equally with neutrals as they do with particles of the same charge, one can write the various correlations as:
\begin{eqnarray}
\gamma'_{\rm ss}&=&\frac{\sum_{i\in +,j<\in +,i\ne j}(p_{xi}p_{xj}-p_{yi}p_{yj})+\sum_{i\in -,j\in -,i\ne j}(p_{xi}p_{xj}-p_{yi}p_{yj})}
{2M_{\pm}(M_\pm-1)\langle p_t^2\rangle},\\
\nonumber
\gamma'_{\rm os}&=&\frac{\sum_{i\in +,j\in -}(p_{xi}p_{xj}-p_{yi}p_{yj})}{M_{\pm}^2\langle p_t^2\rangle}\\
\nonumber
\gamma'_{+0}&=&\frac{\sum_{i\in +,j\in 0}(p_{xi}p_{xj}-p_{yi}p_{yj})}{M_{\pm}M_0\langle p_t^2\rangle}
=\gamma'_{-0}=\gamma'_{\rm ss}.
\end{eqnarray}
Using the symmetries listed above and assuming the multiplicities are large numbers,
\begin{eqnarray}
(M_{\pm}+M_0)\gamma'_{\rm ss}+M_\pm\gamma'_{\rm os}&=&-\frac{1}{2M_{\pm}}\frac{\sum_{i\in \pm}(p_{xi}^2-p_{yi}^2)}{\langle p_t^2\rangle}\\
\nonumber
(M_{\pm}+M_0)\gamma'_{\rm ss}+M_{\pm}\gamma'_{\rm os}&=&-v'_2.
\end{eqnarray}
If $M$ refers to the net multiplicity of charged particles, $M=2M_\pm$, and if $f_0$ is the fraction of particles that are neutral, $M_0=f_0(M+M_0)$, one express the result as:
\begin{equation}
\frac{1}{2}\left(\frac{1+f_0}{1-f_0}\right)\gamma'_{\rm ss}+\frac{1}{2}\gamma'_{\rm os}=-\frac{v'_2}{M}.
\end{equation}
For one third of the particles being neutral, $f_0=1/3$, 
\begin{equation}
\label{eq:psumrule3species}
\gamma'_{\rm ss}+\frac{\gamma'_{\rm os}}{2}=-\frac{v'_2}{M}.
\end{equation}
Since this linear combination of same-sign and opposite-sign contributions seems most connected to the momentum sum rule, we define the quantities
\begin{eqnarray}
\gamma'_{\rm pbal}&\equiv& \gamma'_{\rm ss}+\frac{\gamma'_{\rm os}}{2},\\
\nonumber
\gamma_{\rm pbal}&\equiv& \gamma_{\rm ss}+\frac{\gamma_{\rm os}}{2}.
\end{eqnarray}
By explaining both $\gamma_{\rm qbal}$, which focuses on correlations from charge conservation, and $\gamma_{\rm pbal}$ which is most related to momentum conservation, one would reproduce both the same- and opposite-sign correlations. The advantage of this choice is that the individual quantities can best be interpreted in terms of a single effect.

Following the lines from the previous subsection, one can also account for lack of perfect acceptance, at which point one finds the following final expressions,
\begin{eqnarray}
\label{eq:sumrules}
\frac{1}{2}\left(\frac{1+f_0}{1-f_0}\right)\gamma'_{\rm ss}+\frac{1}{2}\gamma'_{\rm os}&=&-f_P\frac{v_2}{M},\\
\nonumber
\gamma'_{\rm ss}+\frac{\gamma'_{\rm os}}{2}&=&-f_P\frac{v'_2}{M},~{\rm for}~f_0=1/3.
\end{eqnarray}
Here, the multiplicity $M$ is the actual measured multiplicity, and $f_P$ is the fraction of balancing momentum observed given the limited acceptance and efficiency. For a constant efficiency in a limited acceptance, efficiency reduces both $f_P$ and $M$ by the same factor, as expected since correlations should be unchanged by a constant efficiency. Thus, this expression also works if $M$ is the total multiplicity inside the acceptance assuming perfect efficiency, as long as $f_P$ refers to the fraction of momentum inside the acceptance, not just what is measured. For complicated efficiencies and acceptances, the model independence becomes somewhat compromised. However, even in that case the expression suggests that the combination $\gamma'_{\rm ss}+\gamma'_{\rm os}/2$ is a good choice for an observable that would best isolate the effects of momentum conservation. Additionally, one must consider the qualifier that analyses of $v_2$ yields a variety of results, differing by 10-20\%, depending on how the analysis was performed. For the purposes of the sum rule, the value for $v_2$ should include some ``non-flow'' contributions, such as those arising from resonances.

This is a good point for summarizing the assumptions that led to the derivation of Eq. (\ref{eq:sumrules}). First, it was assumed that the momentum transfers responsible for elliptic flow are locally balanced, i.e., not spread over many units of rapidity. Secondly, it was assumed that an unmeasured neutral particle would balance the momentum of a given charged particle identically as another charged particle of the same sign. If observations are not in line with Eq. (\ref{eq:sumrules}), one should question these two assumptions.

\section{Momentum Diffusion and Viscosity}\label{sec:vdiffusion}

It was shown in \cite{GavinPRL,Gavin:2007zz,Gavin:2008ta} that transverse momentum diffuses longitudinally with the viscosity playing a role in determining the diffusion constant. Restating the arguments, one can consider a boost-invariant system with a momentum per rapidity slice, $P_{x}(\eta)$ or $P_{y}(\eta)$, where 
\begin{equation}
P_x(\eta,\tau)=\tau\int dxdy \tilde{T}_{0x}(x,y,\eta,\tau),
\end{equation}
where $\tilde{T}_{\alpha\beta}$ is the stress-energy tensor as viewed by an observer moving with rapidity, $y=\eta$. Here, $\eta$ and $\tau$ play role of the longitudinal position and the proper time in Bjorken coordinates,
\begin{equation}
z=\tau\sinh\eta,~t=\tau\cosh\eta.
\end{equation}
Momentum conservation, $\partial_\tau T_{0x}+\partial_iT_{ix}=0$, yields
\begin{equation}
\partial_\tau P_x(\eta,\tau)=-\frac{1}{\tau}\partial_\eta\int dxdy~\tilde{T}_{xz}.
\end{equation}
Assuming the Navier-Stokes equation,
\begin{equation}
\tilde{T}_{xz}=-\frac{\eta_s}{\tau} \partial_\eta v_x,~~v_x=\frac{\tilde{T}_{0x}}{\epsilon+P},
\end{equation}
where $\eta_s$ is the shear viscosity, one then finds
\begin{equation}
\label{eq:vdiffusion}
\partial_\tau P_x(\eta,\tau)=\frac{\eta_s}{(P+\epsilon)}\frac{1}{\tau^2}\partial^2_\eta P_x(\eta,\tau).
\end{equation}
For small $z$, $(1/\tau)\partial_\eta=\partial_z$, so the diffusion coefficient becomes
\begin{equation}
D=\frac{\eta_s}{P+\epsilon}.
\end{equation}

For the correlations in a boost-invariant system, we define the correlation function for the interval $0<\eta<H$, with the rapidty range $H\rightarrow\infty$,
\begin{eqnarray}
\Pi_x(\Delta\eta)&\equiv&\frac{1}{H}\int_0^{H} d\eta_1d\eta_2~\langle\langle P_x(\eta_1,\tau)P_x(\eta_2,\tau)\rangle\rangle\cdot\delta(\eta_1-\eta_2-\Delta\eta)\\
\nonumber
&=&\frac{1}{H}\left\langle\left\langle\sum_{i,j}p_{xi}p_{xj}\delta(\eta_i-\eta_j-\Delta\eta)\right\rangle\right\rangle,
\end{eqnarray}
where the averaging, $\langle\langle\cdots\rangle\rangle$, denotes an average over events. If one subtracts the $i=j$ terms, one could identify the sum as the numerator used for calculating $\gamma'_{\rm pbal}$. This motivates the definition of $\Pi'_x$,
\begin{eqnarray}
\label{eq:piprimedef}
\Pi'_x(\Delta\eta)&\equiv&\Pi_x(\Delta\eta)-\frac{1}{H}\left\langle\left\langle\sum_i p_{xi}^2\right\rangle\right\rangle\delta(\Delta\eta)\\
\nonumber 
&=&\frac{1}{H}\left\langle\left\langle\sum_{i\ne j}p_{xi}p_{xj}\right\rangle\right\rangle.
\end{eqnarray}
Using Eq. (\ref{eq:vdiffusion}), the evolution of $\Pi'_x$ is
\begin{equation}
\partial_\tau\Pi'_x(\Delta\eta)=\frac{2D}{\tau^2}\partial_\eta^2\Pi'(\Delta\eta),~~\Delta\eta\ne 0.
\end{equation}
The condition that one must stay away from $\Delta\eta=0$ comes from the diffusion equation not being valid for scales so small that the individual particles correlations with themselves does not appear. To include $\Delta\eta=0$, one can consider the effects of collisions on $\Pi'$ defined in Eq. (\ref{eq:piprimedef}). Comparing the instants immediately before and after a collision between particles $i$ and $j$, the contribution to $\Pi'$ involving other particles than those involved in the specific collision, $k\ne i, k\ne j$, does not change because the total momentum carried by $i$ and $j$ does not change. However, the contribution to $\Pi'$ involving the colliding pair, $(i,j)$, does change suddenly. To calculate the change, consider the fact that the net momentum does not change, i.e.,
\begin{equation}
\Delta(p_{ix}+p_{jx})^2=0,
\end{equation}
One can then identify the change in $\Pi'$ as
\begin{eqnarray}
\Delta\Pi'_x(\Delta\eta)&=&\frac{1}{H}\Delta(2p_{ix}p_{jx})\delta(\Delta\eta)\\
\nonumber
&=&-\frac{1}{H}\Delta(p^2_{ix}+p^2_{jx})\delta(\Delta\eta),
\end{eqnarray}
where it is assumed that the two scattering particles have the same coordinate $\eta$. Similar expressions can be derived for any $m\rightarrow n$ process. The rate of change of $\Pi'$ due to the instantaneous collisions thus has two pieces. First, the motion of the particles in between collisions is responsible for the diffusive term. A second term involving the instantaneous changes due to collisions of two particles at the same rapidity can be identified as the collision rate multiplied by the average change of $p_x^2$ per collision,
\begin{equation}
\label{eq:Pidiffusion}
\partial_\tau\Pi'_x(\Delta\eta)=\frac{2D}{\tau^2}\partial_\eta^2\Pi'_x(\Delta\eta)
-\frac{d}{d\tau}\frac{1}{H}\left\langle\left\langle\sum_i p_{xi}^2\right\rangle\right\rangle\delta(\Delta\eta).
\end{equation}

The second term in Eq. (\ref{eq:Pidiffusion}) is a source term for the diffusion. The quantity $\Pi'_x-\Pi'_y$ is proportional to the numerator used to define $\gamma'$, as in Eq. (\ref{eq:gammaprimedef}), if it were binned for pairs of a specific relative momentum. Thus,
\begin{equation}
\gamma'(\Delta\eta,\tau)= (\Pi'_x-\Pi'_y)\frac{H}{M^2\langle p_t^2\rangle}
=\frac{\langle\langle\sum_{\i\ne j}(p_{xi}^2-p_{yi}^2)\delta(\eta_i-\eta_j-\Delta\eta)\rangle\rangle}{M^2\langle p_t^2\rangle},
\end{equation}
where $\langle p_t^2\rangle$ is defined for large asymptotic times.

Since $\gamma'$ is proportional to $\Pi'_x-\Pi'_y$, which obey diffusion equations, so does $\gamma'$,
\begin{equation}
\label{eq:finaldiffusion}
\partial_\tau\gamma'(\Delta\eta,\tau)-\frac{2D}{\tau^2}\partial_\eta^2\gamma'(\Delta\eta,\tau)
=-\frac{1}{M}\delta(\Delta\eta)\frac{d}{d\tau}v'_2(\tau).
\end{equation}
The source term for the diffusion represents the rate at which $v'_2$ rises, defined in Eq. (\ref{eq:v2primedef}), with all the strength located at $\Delta\eta=0$. The source term thus maintains the momentum sum rule for $\gamma'$ derived in Eq. (\ref{eq:v2primedef}). 

To gain an initial understanding of the scale one would expect the diffusion to extend, one can consider a boost-invariant system of multiplity, $dM/d\eta$, in area $A$, of particles colliding with fixed cross sesction $\sigma$. The diffusion constant increases with time due to the fall in density. For massless particles, one can crudely express the viscosity in terms of the cross sections and density as,
\begin{equation}
\eta\approx \frac{1}{5}\frac{P+\epsilon}{n\sigma}.
\end{equation}
If the density is $(dM/d\eta)/(A\tau)$ and if one considers times for which the source term is zero, the diffusion equation becomes
\begin{equation}
\partial_\tau\Gamma'(\Delta\eta)=\frac{2A/5}{(dM/d\eta)\sigma\tau}\partial_{\Delta\eta}^2\Gamma'(\Delta\eta).
\end{equation}
A solution to the equation is
\begin{eqnarray}
\Gamma'(\Delta\eta)&\propto&\frac{\exp{-(\Delta\eta)^2/4ds}}{\sqrt{s}},\\
\nonumber
s&\equiv&\ln\tau,~d\equiv \frac{2A/5}{(dM/d\eta)\sigma},
\end{eqnarray}
where the transverse area is $A$. The width of a diffusive cloud that  began as a delta function at $\tau_{\rm elliptic}$ is,
\begin{equation}
\sigma_\eta=2\sqrt{\frac{A/5}{(dM/d\eta)\sigma}\ln(\tau/\tau_{\rm elliptic})}.
\end{equation}
Thus, $\tau_{\rm elliptic}$ would represent a typical time for which elliptic flow is generated. This tends to be early in the collision, perhaps a few fm/$c$ after thermalization. For central collisions, the transverse area is a few tens of square fm, and the multiplicity can be several hundreds. If the characteristic cross sections are a few tens of mb, and if the ratio $\tau/\tau_{\rm elliptic}$ is of order ten, the spread $\sigma_\eta$ should be roughly a half unit of rapidity. This is the same scale as the spread due to the final thermal motion. The breakup conditions and flow can be well understood by either blast-wave analyses, or by fitting to hydrodynamic or to hybrid hydrodynamic/cascade models. This would allow isolation and determination of $\sigma_\eta$ as a function of centrality. If taken in concert with a detailed model of the evolution that describes when elliptic flow was generated, this might provide insight into the viscosity. 

It should be emphasized that the simple calculations of this subsection, based on a boost-invariant model with no transverse flow, and with an assumption of a fixed number of massless particles with fixed cross sections, certainly cannot be seriously applied to extract properties such as the viscosity from data. However, this simple picture provides insight into the role of viscosity and the importance of understanding the origins of elliptic flow.

\section{Pion Cascade}\label{sec:model}

In order to investigate the ideas put forth in the previous sections, and in order to see whether the ideas might ultimately explain measurements at RHIC, a simple model was constructed in which a gas of pions evolved according to simple s-wave cross scatterings with a fixed cross section. The initial momenta were assigned according to a thermal distribution with a uniform temperature. The space time coordinates were assigned according to the distribution,
\begin{eqnarray}
\frac{dN}{dxdyd\eta d\tau}&\propto& \exp\left\{-\frac{x^2}{2R_x^2}-\frac{y^2}{2R_y^2}-\frac{\eta^2}{2\eta_G^2}\right\}\delta(\tau-\tau_0),\\
\nonumber
\tau&\equiv&\sqrt{t^2-z^2},~~\eta\equiv \tanh^{-1}(z/t),
\end{eqnarray}
with the collective velocity in the $z$-direction given by $v_z=z/t$. These are known as Bjorken coordinates, which are invariant to boosts in the longitudinal direction, similar to a flat-space Hubble expansion, in the limit $\eta_G\rightarrow\infty$. After picking source points according to the distribution above, two-particles were generated according to the thermal distribution for each point, then boosted according to the longitudinal collective velocity. Two thirds of the pairs were chosen to be $\pi^+\pi^-$ pairs, and one third as $\pi^0\pi^0$ pairs. This enforced local charge conservation at the extreme in the initial state, which then relaxed as the system evolved. The collective nature, short mean free paths, led to strong correlations between the balancing particles in both azimuthal angle $\phi$ and in rapidity. The evolution involved straight-line trajectories punctuated by s-wave collisions with fixed cross sections that were independent of isospin. All the simulations  used values of $T_0=250$ MeV for the initial temperature, $\tau_0=1.0$ fm/$c$ for the initial time and $\eta_G=2$. The fixed cross section $\sigma$ was set at 15 or 30 mb to test sensitivity of the final-state correlations to diffusion.  Runs were made with 1000 particles and with initial transverse radii of both $R_x=1.5, R_y=3.0$ fm, and $R_x=1.8, R_y=2.5$ fm. These two geometries give the same central densities, thus making it possible to test the scaling with anisotropies. Additionally, runs were made with a smaller system, 250 particles, with radii of $R_x=0.9, R_y=1.25$ fm, in order to vary the size while keeping the density fixed. Due to the high initial pion density, $\sim 1$ pions/fm$^3$, and the large cross sections, $\sigma=30$ mb$~=3$ fm$^2$, momentum transfer can diffuse at super-luminous speeds \cite{Kortemeyer:1995di,Cheng:2001dz}. To prevent such unphysical behavior the particles were oversampled by a factor $N_{\rm sample}$ combined with a reduction of the cross section by a factor $1/N_{\rm sample}$. In the limit $N_{\rm sample}\rightarrow \infty$, super-luminous behavior would disappear while maintaing roughly the same mean free path as for $N_{\rm sample}=1$. For the 15 mb case, it was found that $N_{\rm sample}=4$ was sufficient to approach the $N_{\rm sample\rightarrow\infty}$ limit, while $N_{\rm sample}=8$ was used for the $\sigma=30$ mb calculations. Analyses were made for two acceptances: a perfect acceptance and one that crudely mimics STAR's acceptance. The STAR acceptance was modeled by considering only those particles with pseudo-rapidities between -1 and 1, and that had transverse momenta between 200 MeV/$c$ and 2 GeV/$c$.

\begin{table}
\begin{center}
\begin{tabular}{|c|c|c|c|c|c|}\hline
acceptance & $\sigma$ (mb)& $R_x,R_y$(fm)& $v_2$ & $\gamma_{\rm pbal}M/v_2$ & $\gamma'_{\rm pbal}M/v'_2$
\\ \hline\hline
FULL & 15 & 1.8, 2.5 & 0.0677 & -1.901 & -0.985
\\ \hline
FULL & 30 & 1.8, 2.5 & 0.0776 & -2.034 & -1.060
\\ \hline\hline
FULL & 15 & 1.5, 3 & 0.137 & -2.075 & -1.007
\\ \hline
FULL & 30 & 1.5, 3 & 0.158 & -2.051 & -1.002
\\ \hline\hline
FULL & 15 & 0.9, 1.25 & 0.0503 & -1.798 & -0.982
\\ \hline
FULL & 30 & 0.9, 1.25 & 0.0562 & -1.959 & -0.983
\\ \hline\hline
STAR & 15 & 1.8, 2.5 & 0.0910 & -1.324 & -0.817
\\ \hline
STAR & 30 & 1.8, 2.5 & 0.0997 & -1.307 & -0.779
\\ \hline\hline
STAR & 15 & 1.5, 3 & 0.186 & -1.282 & -0.789
\\ \hline
STAR & 30 & 1.5, 3 & 0.204 & -1.407 & -0.847
\\ \hline\hline
STAR & 15 & 0.9, 1.25 & 0.0667 & -1.1354 & -0.763
\\ \hline
STAR & 30 & 0.9, 1.25 & 0.0697 & -1.2854 & -0.833
\\ \hline
\end{tabular}
\caption{\label{table:gammap}The results for $v_2$ and the scaled correlations $M\gamma_{\rm pbal}/v_2$ and $M\gamma'_{\rm pbal}/v'_2$ are shown for different acceptances, cross sections, initial radii and sizes. The scaling (multiplied by the multiplicity/$v_2$) was chosen so that $\gamma'_{\rm pbal}M/v'_2$ would be unity for full acceptance according to the sum rule, Eq. (\ref{eq:psumrule3species}). For the calculations with STAR acceptance, the values of $\gamma'$ hovered around 0.8, indicating that for any given observed particle about 80\% of the balancing momentum responsible for elliptic flow would be found within $-1<\eta<1$. The scaled values of $\gamma$ were consistently nearly twice as large as the values for $\gamma'$. Given the statistical errors, of the order of 10\% using 40,000 events, it is difficult to see any sensitivity of the scaled moments to doubling the cross section, doubling the system size, or roughly doubling the anisotropy.
Given the consistency of the $\gamma'_{\rm pbal}$ and $\gamma_{\rm pbal}$ values, these calculations would lead one to expect similar behavior in STAR analysis, i.e., the scaled values of $\gamma$ would stay roughly constant with centrality, and within a few tens of percent of 1.3.}
\end{center}
\end{table}

Results for $M\gamma'_{\rm pbal}/v'_2$ and $M\gamma_{\rm pbal}/v_2$ are shown in Table \ref{table:gammap} for both the case of perfect acceptance and for STAR's acceptance, for both anisotropies, for both cross sections, and for the scaled-down size. The multiplicity $M$ is the average number of charged particles with pseudo-rapidities between -1 and 1. The statistical accuracy of the quantities are of the order of 5-10\%. Calculations with smaller $v_2$, or with smaller acceptances, tended to have larger uncertainties. For that reason, runs were not made with smaller anisotropies.

The momentum sum rule applies only to the case of full acceptance, and only for the primed quantities. The results in Table \ref{table:gammap} show that finite acceptance reduces the ratios, $\gamma'_{\rm pbal}M/v'_2$, by approximately 20\%. This reduction implies that for each particle within the specified rapidity range, roughly 80\% of the balancing transverse momentum exchanges responsible for elliptic flow can also be found within the rapidity range. The reduction appears fairly uniform for all the cases studied. Unfortunately, STAR did not report the primed ($p_t$ weighted) quantities in \cite{:2009txa}, but instead reported unprimed quantities. Fortunately, the unprimed quantities in  Table \ref{table:gammap} are also fairly constant. The steadiness of the values would lead one to expect values for $\gamma_{\rm pbal}M/v_2$ to be near $1.3$, and to vary rather little with centrality. Since the ratio depends on the fraction of balancing momentum found within the acceptance, and since that fraction might well depend on centrality, it would not be surprising if the  ratio varied by tens of percent.

\begin{figure}
\centerline{\includegraphics[width=0.45\textwidth]{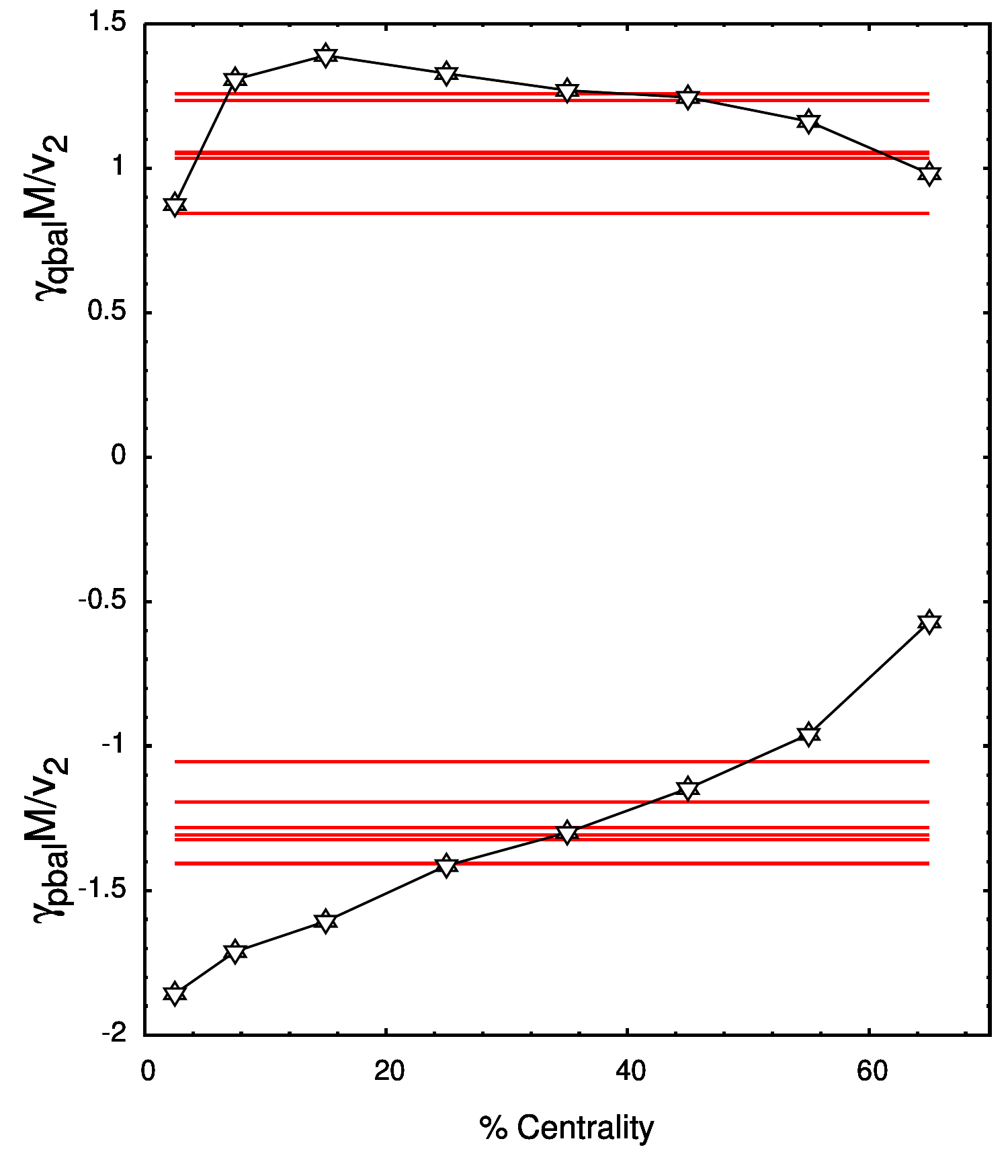}}
\caption{\label{fig:gvscent}(color online)
Results from STAR are shown for the ratios $M\gamma_{\rm pbal}/v_2$ and for $M\gamma_{\rm qbal}/v_2$. The ratios $M\gamma_{\rm qbal}/v_2$, which are related to momentum conservation, vary by a factor of 3 from the most peripheral to the most central collisions. In contrast, the values found with the cascade model (horizontal lines) vary only by 10\%. The parameters for each of these values is described in detail in Table \ref{table:gammap}, with the variation being statistically consistent with being constant, even though geometries, sizes and cross sections were all varied by a factor of two for different runs. The ratio $M\gamma_{\rm qbal}/v_2$ is identified with charge conservation. Despite the fact that the model poorly matches spectra or flow variables, the values found with the model were within a few tens of percent of the ratio found experimentally.}
\end{figure}
Experimental results for the ratio, $\gamma_{\rm pbal}M/v_2$, are displayed in Fig. \ref{fig:gvscent}. The expectations described above are not well validated by the experimental observations. Although the sign and overall magnitude are of the right order, the experimental ratio increased by a factor of three from peripheral to central collisions, whereas the ratios extracted from model varied rather little from $1.3$, roughly independent of changes in the initial anisotropy, the cross-section, or the system size. This discrepancy between the model and experiment at the qualitative level suggest a significant shortcoming in the model.

Aside from the large centrality dependence, a second surprise is in the size of the ratio for the most central collisions. For ratios much smaller than 1.3, a simple explanation would involve a larger fraction of the balancing momentum being carried outside the acceptance. For ratios much higher than 1.3, one would look for reasons to expect a larger fraction of the momentum being balanced by particles inside the acceptance than what was occurring in the pion cascade. However, for the pion cascade $\approx 80$\% of the momentum sum rule was being realized, which gives little room for overshooting the ratios from the pion cascade. Another possibility would be that a larger fraction of the balancing momentum was being carried by neutrals in the real collisions as compared to the presumed $1/3$ of the pion cascade. Yet another possibility might involve the ratio between $\gamma$ and $\gamma'$ in Table \ref{table:gammap}. This ratio might change in realistic models, though it is difficult to foresee by what amount. The better way to handle this would for the experiment to evaluate $\gamma'_{\rm pbal}$ instead, as it would reduce the model dependence involved in interpreting the result. The experimental ratio in Fig. \ref{fig:gvscent} was approximately 40\% higher for the most central collisions than in any of the cascade calculations. Systematic errors for both $v_2$ and for the angular correlations were well over 10\%, so the overshoot might disappear with a more detailed analysis or with higher statistics, but for the moment seems puzzling.

\begin{figure}
\centerline{\includegraphics[width=0.6\textwidth]{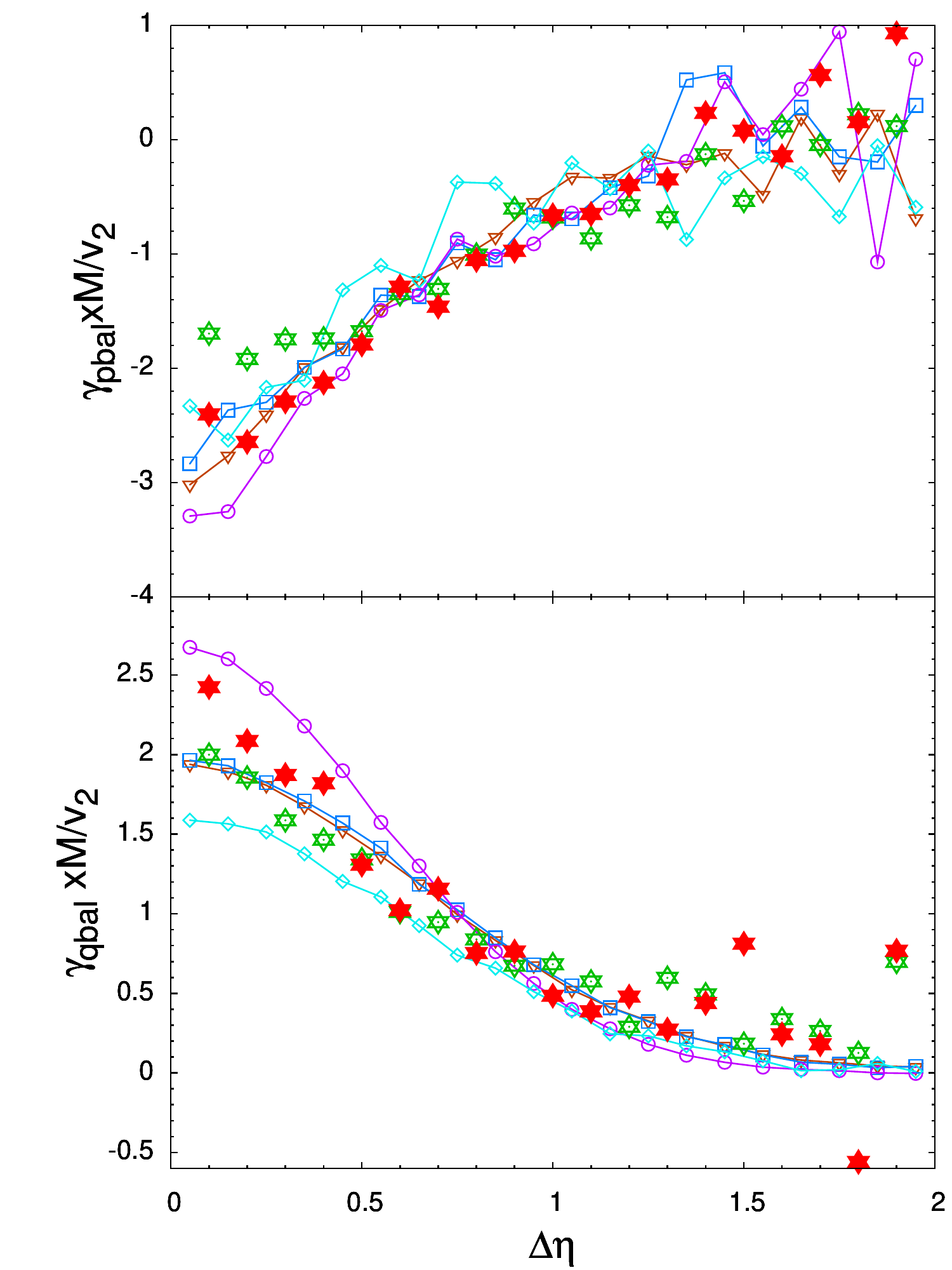}}
\caption{\label{fig:gvsdely}(color online)
Values of the scaled correlation, $M\gamma_{\rm pbal}/v_2$, as determined by STAR, are shown as a function of relative pseudo-rapidity. Results are displayed for two centralities, 10-30\% (red filled stars), and 30-50\% (green open stars). The experimental correlations appear broader for the more peripheral collisions. Also shown are results from four of the cascade simulations: (blue squares) default results with initial sizes of $R_x=1.8, R_y=2.5$ fm and 15 mb cross sections, (purple circles) same as default but with double the cross section, (cyan diamonds) same as default but with half the size for $R_x$ and $R_y$ and with 1/4 the multiplicity, (brown downward triangles) same as default but with a larger anisotropy, $R_x=1.5, R_y=3.0$ fm. The model roughly reproduces the size and extent of the correlations. However, the cascade calculations show little sensitivity to any of the changes, at least within the statistical accuracy of the calculations. Although the statistical significance is marginal, it appears that the fall of $\gamma_{\rm pbal}$ for peripheral collisions seen in STAR's results, plotted in Fig. \ref{fig:gvscent}, derives from an increased broadening.\\
The scaled moments $M\gamma_{\rm qbal}/v_2$, which are associated with charge conservaton, are also displayed. They also have a similar width and height as the experiment. Unlike the scaled values of $\gamma_{\rm pbal}$, the width for these correlations are sensitive to the cross section in the cascade model.}
\end{figure}
STAR has also presented more differential results for the moments displayed in Fig. \ref{fig:gvscent}. In Fig. \ref{fig:gvsdely}, results are shown for the scaled moments, $\gamma_{\rm pbal}M/v_2$ binned as a function of pseudo-relative rapidity. Analyses from STAR for centralities of 10-30\% and for 30-50\% are plotted along side pion cascade results for the initial anisotropy $R_x=1.8, R_y=2.5$ fm, and for two cross sections. As can be seen in Fig. \ref{fig:gvsdely}, the width of the correlation in reasonably well reproduced by the cascade. This centrality range corresponds to the values for which the net moments were reasonably well reproduced by the cascade as shown in Fig. \ref{fig:gvscent}. For the experimental results, it does appear that the correlations are more spread out in relative rapidity for less central collisions, but it would be useful to observe the correlations for both the most central collisions, and for the more peripheral ones, where the net moments are not in line with the expectations of the pion cascade. Performing the more differential analysis in relative pseudo-rapidity for a broader range of centralities could validate whether the dependence of $\gamma_{\rm pbal}$ with centrality indeed comes from a broadening of the correlation for peripheral collisions. Since the sum-rule seems rather robust from a theoretical perspective, it seems likely that the weakening comes from a broadening, but the experimental evidence for this statement needs to be strengthened.

For the moment, we will assume the preliminary interpretation above is true: that the momentum balance responsible for the generation of $v_2$ is local for central collisions, but much broader for peripheral collisions. This behavior is not matched in the pion cascade model, and given that the effects of momentum diffusion increase only logarithmically with time as shown in Sec. \ref{sec:vdiffusion}, it would seem most unlikely for momentum diffusion to substantially increase the width of the correlations in Fig. \ref{fig:gvsdely} for peripheral collisions, unless the processes responsible for elliptic flow occurred within the first few tenths of a fm/$c$. At such early times it would be difficult for elliptic flow to be generated hydrodynamically, even if hydrodynamics were valid, since the characteristic times for generating flow is set by the transverse density gradients. Another possibility is that the assumption that the source for $\gamma'$ in the diffusion equation, Eq. \ref{eq:finaldiffusion}, is not really a delta function. In the summary we discuss the possibility that transverse momentum conservation related to elliptic flow is not purely local in rapidity.

\section{Charge Balance Correlations from the Pion Cascade}

The correlation, $\gamma_{\rm qbal}\equiv\gamma_{\rm ss}-\gamma_{\rm os}$, is largely dominated by charge balance effects. Such correlations are driven by local charge conservation overlaid onto collective flow. These correlations can be well reproduced with a blast-wave calculation tuned to reproduce elliptic flow and spectra if the thermally generated particles are always produced in small neutral ensembles instead of singly \cite{soeren_bf}. Not only do such calculations give reasonable values of $\gamma_{\rm qbal}$, but they are remarkably successful at matching more differential observables such as charge balance functions. Since the pion cascade model presented here is not well tuned to reproduce flow observables or spectra, one should not take the success, or lack thereof, in reproducing data too seriously. Nonetheless, for the sake of completeness, we also show experimental results for $\gamma_{\rm qbal}$ in Fig. \ref{fig:gvscent} along with the values coming from the pion cascade model, which are also listed in Table \ref{table:gammab}. Since non-zero $\gamma_{\rm qbal}$ also requires elliptic flow, and since the effects fall inversely with the multiplicity, the scaled values, $\gamma_{\rm qbal}M/v_2$, are presented as opposed to $\gamma_{\rm qbal}$ itself.
\begin{table}
\begin{center}
\begin{tabular}{|c|c|c|c|c|c|}\hline
acceptance & $\sigma$ (mb)& $R_x,R_y$(fm)& $v_2$ & $\gamma_{\rm qbal}M/v_2$
\\ \hline\hline
FULL & 15 & 1.8, 2.5 & 0.0677 & 1.351
\\ \hline
FULL & 30 & 1.8, 2.5 & 0.0776 & 1.546
\\ \hline\hline
FULL & 15 & 1.5, 3 & 0.137 &  1.326
\\ \hline
FULL & 30 & 1.5, 3 & 0.158 & 1.524
\\ \hline\hline
FULL & 15 & 0.9, 1.25 & 0.0503 & 0.9816
\\ \hline
FULL & 30 & 0.9, 1.25 & 0.0562 & 1.220
\\ \hline\hline
STAR & 15 & 1.8, 2.5 & 0.0910 & 1.056
\\ \hline
STAR & 30 & 1.8, 2.5 & 0.0997 & 1.257
\\ \hline\hline
STAR & 15 & 1.5, 3 & 0.186 & 1.035
\\ \hline
STAR & 30 & 1.5, 3 & 0.204 & 1.235
\\ \hline\hline
STAR & 15 & 0.9, 1.25 & 0.0667 & 0.844
\\ \hline
STAR & 30 & 0.9, 1.25 & 0.0697 & 1.051
\\ \hline
\end{tabular}
\caption{\label{table:gammab}Results from the pion cascade model for the scaled correlations $\gamma_{\rm qbal}$ are shown for different acceptances, cross sections and initial radii. These correlations are driven by charge conservation.}
\end{center}
\end{table}
The experimental values of $\gamma_{\rm qbal}M/v_2$ in Fig. \ref{fig:gvscent} vary between 1.0 and 1.4, whereas the result for the pion cascade tend to be modestly lower. Unlike the scaled values of $\gamma_{\rm pbal}$, there are no sum-rules that apply for $\gamma_{\rm qbal}$. As expected, the scaled values of $\gamma_{\rm qbal}$ are higher for cascade calculations with higher cross sections, or for larger systems. In such cases, balancing charges will by more strongly focused into the same direction. The cascade calculation appears to modestly under-predict the data, perhaps by $\sim 20\%$. This might suggest that in the real collision that many balancing charges are made later in the reaction than 1 fm/$c$, which would lead to the charges being even more tightly correlated. However, that conclusion is premature given the numerous shortcomings of the model. The only firm conclusion to be made here is that the sign and magnitude of $\gamma_{\rm qbal}$ from the model and from data are fairly similar.

Since the cascade model employs isospin-independent cross section, the charge balance correlations can be calculated by comparing only a particle with its balancing partner, thus avoiding the noisy statistical subtraction involving charges from different balancing pairs. Thus, the statistical accuracy for the calculation of $\gamma_{\rm qbal}$ is much better than that for $\gamma_{\rm pbal}$. More quantitative insight into the data can better be ascertained with models that match both the spectra, yields and flow variables, such as what was studied in \cite{soeren_bf}.

\section{Discussion, Summary and Outlook}

As predicated in the previous section, if observations are not in line with the predictions of the model, one should reconsider the two basic assumptions used to derive the sum rule, $\gamma'_{\rm pbal}M/v'_2\rightarrow 1$, for a broad acceptance, or the expectation that $\gamma_{\rm pbal}\approx 1.3$ for STAR's acceptance and only weakly dependent on centrality:
\begin{enumerate}
\item It was assumed that for a charged particle, finding the balancing momentum related to elliptic flow in other particles of the same charge would be identical to finding the balancing momentum in neutral particles. This assumption was explicitly enforced in the pion cascade, but could be different in a more realistic picture. For instance, for positive pions coming from $\rho^+$ decays, one would have a contribution in the same direction for neutral pions, which would not exist between the positive pion and other positive pions. The negative correlation with same-sign pions would then be stronger to compensate.
\item The momentum transfers that build elliptic flow are locally balanced, i.e., if a particle scatters into the reaction plane, the balancing momentum is likely to be found nearby in rapidity. For the momentum to be balanced by particles far removed in rapidity, the transfers need to have occurred early in time, or perhaps even during the passing of the initial nuclei. The speed for which fluctuations of the transverse momentum can traverse the matter in the longitudinal direction is limited by the viscosity in a hydrodynamic context\cite{GavinPRL,Gavin:2007zz,Gavin:2008ta}. 
\end{enumerate}

Violations of (1) would lead to stronger correlations than those predicted in the model. However, it is difficult to expect that by correcting the model with more realistic resonance contributions one might find corrections more than a few tens of percent. Thus, such considerations might help explain why STAR's observations overshot the pion gas by $\approx$40\% for the most central collisions in Fig. \ref{fig:gvscent}, but are unlikely candidates for explaining the factor of three variation in $M\gamma_{\rm qbal}/v_2$ observed by STAR and displayed in Fig. \ref{fig:gvscent}.

Violations of (2) should lead to smaller values for $\gamma_{\rm pbal}$ for a finite acceptance. For the pion cascade calculations, the particles were initialized with zero transverse flow and with symmetric stress energy tensors. The width of the resulting correlations was remarkably insensitive to the cross section employed in the cascade. A shorter mean free path gives a smaller viscosity and thus a smaller diffusion of transverse momentum fluctuations, but also keeps the matter together longer. The surest way to increase the width of the correlations in relative pseudo-rapidity would be to begin the cascade earlier. From the discussion of the diffusion of momentum in Sec. \ref{sec:vdiffusion}, an earlier start time would broaden the final width in the coordinate $\eta$ logarithmically with the starting time. The solutions to the diffusion equation would sugggest that to double the width, one would need to start the cascade at times near 0.1 fm/$c$. These would be well below the times that would be reasonable for modeling the evolution as a cascade, even with partons. Further, one would need to explain how elliptic flow was being generated at such short times as the characteristic time should be set by the transverse size of the collision region divided by the speed of sound.

It is difficult to judge how much the ratio $\gamma_{\rm pbal}/\gamma'_{\rm pbal}$ might change with a more realistic model. Such a model would have a softer equation of state, and would incorporate resonant decays. In order to change the ratio $\gamma_{\rm pbal}/\gamma_{\rm pbal}'$, one would have to spread the momentum balance around as a function of $p_t$. Since $\gamma'_{\rm pbal}$ has the more direct relation to the sum rule, it is less model-dependent. STAR certainly has the ability to determine $\gamma'_{\rm pbal}$, and has even analyzed the correlation differentially in terms of $p_t$. 

Two main conclusions come out of the studies here. First, momentum conservation does seem to play a principle role in determining the combination of opposite- and same-sign correlations, $\gamma_{\rm pbal}$ and $\gamma'_{\rm pbal}$. For a perfect acceptance, the rule seems unquestioned, except for the uncertainty in accounting for the fraction of momentum balance from unobserved particles (mainly neutral particles). Although the centrality dependence was not matched by the pion cascade model, the size of the effect was very much of the right order, and for more central collisions the spread in $\Delta\eta$ was rather well reproduced.

The second conclusion concerns the failure of the pion cascade to reproduce the strong centrality dependence of $\gamma_{\rm pbal}$ seen in the data. STAR's results suggest that the correlation is spread over a large range in rapidity for peripheral collisions. One intriguing explanation of the model's failure is that whereas the model evolution began at 1 fm/$c$, for peripheral collisions it is possible that the elliptic flow was generated while the system was in a coherent state. If coherent structures, like color flux tubes or the color glass condensate, extend a unit of rapidity or more, momentum conservation might no longer be local (assumption \#1 above) in $\eta$. For this explanation to hold, much of the elliptic flow would have to be generated while the system were in such a coherent state. Whereas 1 fm/$c$ represents a small fraction of the time over which flow develops in a central collision, the first fm/$c$ should be more important for peripheral collisions, where the initial logarithmic density gradients are higher, and where the generation of flow finishes earlier. An even more speculative possibility is that in peripheral collisions elliptic flow might be generated at the initial point of contact between the nuclei. This might come from a preference for the individual nucleon-nucleon collisions to have their reaction planes coincide with the $AA$ reaction plane, which could lead to initially anisotropic stress energy tensors with the momentum balance spread over multiple units of rapidity. However, our own quick analysis of Glauber models does not suggest such an effect exists. If early flow is indeed dominating the explanation, it would serve as motivation to reconsider some analyses of elliptic flow, based on hydrodynamic pictures with little initial flow. These analyses have often been used to determine the viscosity of the quark-gluon plasma \cite{Romatschke:2007mq}.

Both STAR's experimental analysis and the model analysis presented here can be improved. The pion cascade model was adequate for illustrating the effects, but is too far from reality to make any serious quantitative conclusion from a comparison to data. The experimental analysis will be strengthened by better statistics, by a more detailed study of systematics (such as determining which method for $v_2$ is more appropriate for scaling $\gamma_{\rm pbal}$) and by evaluating $\gamma'_{\rm pbal}$ in addition to $\gamma_{\rm pbal}$. It is difficult to determine how to fix many of the deficiencies of the model. More realistic models require hydrodynamic prescriptions for the intermediate stage as well as dynamic models of the field-dominated pre-equilibrium stages, thus incorporating the generation and propagation of momentum fluctuations in such models is necessary. Such advances are clearly challenging, but understanding the questions raised in the previous paragraphs would be valuable. This class of observables might ultimately provide insight into the state and nature of the pre-thermalized state, or might even assist in constraining bulk properties of quark-gluon plasma. 

\acknowledgments{This work was supported by the U.S. Department of Energy, Grant No. DE-FG02-03ER41259. The authors also thank Denes Molnar and Sergei Voloshin for generously sharing their insight.}


\begin{thebibliography}{99}

\bibitem{:2009uh}
  B.~I.~Abelev {\it et al.}  [STAR Collaboration],
  Phys.\ Rev.\ Lett.\  {\bf 103}, 251601 (2009)
  [arXiv:0909.1739 [nucl-ex]].
 
\bibitem{:2009txa}
  B.~I.~Abelev {\it et al.}  [STAR Collaboration],
  Phys.\ Rev.\  C {\bf 81}, 054908 (2010)
  [arXiv:0909.1717 [nucl-ex]].

\bibitem{Kharzeev:2004ey}
  D.~Kharzeev,
  Phys.\ Lett.\  B {\bf 633}, 260 (2006)
  [arXiv:hep-ph/0406125].
  
\bibitem{Kharzeev:2009mf}
  D.~E.~Kharzeev,
  Nucl.\ Phys.\  A {\bf 830}, 543C (2009)
  [arXiv:0908.0314 [hep-ph]].

\bibitem{Bzdak:2009fc}
  A.~Bzdak, V.~Koch and J.~Liao,
  Phys.\ Rev.\  C {\bf 81}, 031901 (2010)
  [arXiv:0912.5050 [nucl-th]].

\bibitem{Wang:2009kd}
  F.~Wang,
  Phys.\ Rev.\  C {\bf 81}, 064902 (2010)
  [arXiv:0911.1482 [nucl-ex]].

\bibitem{soeren_bf}
S. Schlichting and S. Pratt, in preparation, 2010

\bibitem{Schlichting:2010na}
  arXiv:1005.5341 [nucl-th].

\bibitem{Aggarwal:2010ya}
  M.~M.~Aggarwal {\it et al.}  [STAR Collaboration],
  Phys.\ Rev.\  C {\bf 82}, 024905 (2010)
  [arXiv:1005.2307 [nucl-ex]].

\bibitem{Pratt:2010gy}
  S.~Pratt,
  arXiv:1002.1758 [nucl-th].
	
\bibitem{Bzdak:2010fd}
  A.~Bzdak, V.~Koch and J.~Liao,
  arXiv:1008.4919 [nucl-th].

\bibitem{GavinPRL} S. Gavin and M. Abdel-Aziz, Phys. Rev. Lett. {\bf 97}, 162302 (2006).

\bibitem{Gavin:2008ta}
  S.~Gavin and G.~Moschelli,
  J.\ Phys.\ G {\bf 35}, 104084 (2008)
  [arXiv:0806.4366 [nucl-th]].
  
\bibitem{Gavin:2007zz}
  S.~Gavin and M.~Abdel-Aziz,
  Braz.\ J.\ Phys.\  {\bf 37}, 1023 (2007).


\bibitem{Kortemeyer:1995di}
  G.~Kortemeyer, W.~Bauer, K.~Haglin, J.~Murray and S.~Pratt,
  Phys.\ Rev.\  C {\bf 52}, 2714 (1995)
  [arXiv:nucl-th/9509013].
  
\bibitem{Cheng:2001dz}
  S.~Cheng {\it et al.},
  Phys.\ Rev.\  C {\bf 65}, 024901 (2002)
  [arXiv:nucl-th/0107001].
  
\bibitem{Romatschke:2007mq}
  P.~Romatschke and U.~Romatschke,
  Phys.\ Rev.\ Lett.\  {\bf 99}, 172301 (2007)
  [arXiv:0706.1522 [nucl-th]].
  
\end{thebibliography}
\end{document}